\documentclass[aps,prl,twocolumn,superscriptaddress]{revtex4}
\usepackage[bookmarks = false,colorlinks = true,linkcolor = blue,urlcolor= black,citecolor = blue]{hyperref}
\usepackage{graphicx}
\usepackage{dcolumn}
\usepackage{bm}
\usepackage{amsmath}
\usepackage{amssymb}
\usepackage{multirow}
\allowdisplaybreaks

\def\tr{{\rm tr}}

\def\aff{{\rm aff}}
\def\conv{{\rm conv}}

\def\C{\mathcal{C}}
\def\I{\mathcal{I}}
\def\0{\vec{0}}
\def\u{\vec{u}}
\def\v{\vec{v}}
\def\w{\vec{w}}
\def\vg{\vec{g}\,}
\def\c{\vec{c}\,}
\def\co{\vec{c}\,{}^{\bullet}}
\def\cw{\vec{c}\,{}^{\circ}}

\def\g{\mathrm{guess}}
\def\trho{\widetilde{\rho}}
\def\bea{\begin{eqnarray}}
\def\eea{\end{eqnarray}}

\newtheorem{theorem}{Theorem}

\begin{document}

\title{Complete analysis to minimum-error discrimination of mixed four qubit states with arbitrary prior probabilities}

\author{Donghoon Ha}
\affiliation{Department of Applied Mathematics and Institute of Natural Sciences, Kyung Hee University, Yongin 17104, Republic of Korea}
\author{Younghun Kwon}
\email{yyhkwon@hanyang.ac.kr}
\affiliation{Department of Applied Physics, Center for Bionano Intelligence Education and Research, Hanyang University, Ansan 15588, Republic of Korea}

\date{\today}

\begin{abstract}
In this work, we provide a complete analysis to minimum-error discrimination of mixed  
four qubit states with arbitrary prior probabilities.
For the complete anaysis, the most important work to do is to find the
necessary and sufficient conditions for the existence of null measurement operator. 
From the geometric structure of qubit states, we obtain the analytic
condition for deciding the existence of a null operator in
minimum-error measurement for mixed four qubit states, which also gives the 
necessary and sufficient conditions for every optimal POVM to have non-zero elements. Using the condition, we completely analyze minimum-error discrimination of mixed four
qubit states with arbitrary prior probabilities. 
\end{abstract}

\maketitle

\indent Quantum state discrimination\cite{ref:chef1,ref:barn1,ref:berg1} is one of the most fundamental tasks in quantum information theory. It identifies the fundamental limit of
distinguishability of systems dictated by postulates of quantum
theory, and is also useful in various quantum information
applications\cite{ref:bae1}. When one of states $\{\rho_{i}\}_{i=1}^N$ is prepared with {\it a
priori} probability $q_i$ from $\{ q_{i} \}_{i=1}^N$,  we may denote the situation as $\{ q_i ,\rho_i\}_{i=1}^N$.
Then quantum state discrimination is the process of verifying the unknown state out of the
possibilities of $\{ \rho_i\}_{i=1}^N$. \\
\indent Minimum-error discrimination (MD)\cite{ref:hels1,ref:hole1,ref:yuen1} means
the process achieved by minimising the average error or equivalently
maximising probability of making correct guesses in average. The
latter is called the guessing probability. This seeks optimization
of measurement, a set of positive-operator-valued-measures (POVMs),
$\{ M_i\}_{i=1}^N$, which is a non-negative resolution of the
identity operator i.e. $ \sum_{i} M_i = I $ with $M_i \geq 0$. Then,
the guessing probability is expressed as,
\begin{equation}\label{eq:guessingprob}
p_{\g} = \max_{\{ M_i\}_{i=1}^N} \sum_{i=1}^{N} q_i \,p( i| i )
\end{equation}
where $p(i|j) = \tr[\rho_j M_i]$ denotes the probability of
obtaining outcome $i$ when $\rho_j$ has been prepared. So far,
two-state minimum error discrimination, regardless of the dimension of quantum state, is the case that optimal discrimination is
completely solved, for arbitrary states and prior
probabilities, in an analytic form\cite{ref:hels1,ref:herz1}. Apart
from the case, despite the fundamental and practical importance,
little has been known as analytic formula for the
guessing probability or also for finding optimal measurement\cite{ref:ban1,ref:elda1,ref:chou1,ref:moch1,ref:sams1,ref:bae2,ref:bae3,ref:ha1,ref:ha2, ref:ha3}.\\
\indent Recently, much effort has been devoted to deriving analytical methods for
optimal discrimination of qubit states.
One of the approaches\cite{ref:deco1,ref:bae2,ref:bae3,ref:ha1,ref:ha2} is to make use of semidefinite programming\cite{ref:elda2,ref:boyd1}
together with the Bloch sphere expression of qubit states. In fact, for
a non-negative operator $A$, its Bloch vector is denoted by
$\vec{r}_A$ such that
\begin{equation}
A = \frac{\tr[A]}{2}(I+ \vec{r}_A \cdot \vec{\sigma}),
\nonumber
\end{equation}
where $\vec{\sigma}$ is the Pauli matrices $(\sigma_X, \sigma_Y, \sigma_Z)$.\\
\indent An optimization task in semidefinite programming can be generally
formulated in the so-called linear complementarity program (LCP)\cite{ref:bae1,ref:bae3}
that generalizes primal and dual problems. The LCP approach to
optimal qubit state discrimination has also been formulated and
presented in another form of geometric method. The LCP of qubit
state discrimination is to use operators $K$ and $\{r_i , \trho_{i} \}_{i=1}^N$ such that for
all $i=1,\cdots,N$
\begin{equation}\label{eq:optc}
K = q_i \rho_i + r_i \trho_i
,~~\mathrm{and} ~~ r_i \tr[\trho_i M_i ] = 0,
\end{equation}
where $\{r_i \}_{i=1}^N$ are constants, not probabilities, and $\{
M_i\}_{i=1}^N$ is a measurement. Once these operators $K$, $\{r_i ,
\trho_{i} \}_{i=1}^N$, and $\{M_i \}_{i=1}^N$ satisfying the above
are found, they define optimal operators for optimal discrimination.
Then the guessing probability corresponds to $\tr[K]$ as it is
shown in Eq.~\eqref{eq:guessingdual}. 
\begin{equation}\label{eq:guessingdual}
p_{\g} = \min_{K\geq q_i \rho_i \,\forall i} \tr[K].
\end{equation}
A set of POVMs satisfying the
orthogonality with $\{ \trho_i \}_{i=1}^N$ forms an optimal
measurement. As the LCP approach deals with a list of equations that
characterizes optimal parameters of the primal and dual problems, it
could be considered more complicated in general. The advantage lies
at the fact that the structure of the problem is exploited. In fact,
the optimality conditions in Eq.~\eqref{eq:optc} can find an
underlying geometry of states and the optimal parameters in the
state space.\\
\indent If prior probabilities are not equal, the
LCP approach has not yet found a geometric method for solving
optimal discrimination. However, remarkably, the optimization was successfully achieved in three mixed qubit states with arbitrary prior probabilities\cite{ref:ha1}.
A closed formula for three mixed qubit states could be derived. 
Note that it is a formula for any three qubit states given with arbitrary prior 
probabilities. As far as we are aware of, it is the only analytic
form of optimal state discrimination other than the two-state
discrimination when no prior information is assumed on given states
but the dimension
$d=2$.\\
\indent However, minimum error discrimination of four mixed qubit states with arbitrary prior probabilities is still an unsloved problem. 
And in this work, we will provide a complete analysis to minimum error discrimination of four mixed qubit states with arbitrary prior probabilities.
For minimum error discrimination of four mixed qubit states with arbitrary prior probabilities, we will analyze the
necessary and sufficient conditions for the existence of null measurement operator. Specifically, from the geometric structure of qubit states, we will obtain the analytic
condition for deciding the existence of a null operator in
minimum-error measurement for four qubit states. It equips us with the 
necessary and sufficient conditions for every optimal POVM to have non-zero elements. 
From the condition, we can completely analyze minimum-error discrimination of four mixed 
qubit states with arbitrary prior probabilities.\\
\indent The purpose of MD is to minimize the guessing error, so that
guessing the quantum state produced is done as accurately as
possible. The dual problem to minimize $\rm \tr \it [K]$ in the
condition $K =q_{i} \rho_{i}+r_{i} \trho_{i}(\forall i)$ can
be obtained by introducing the Lagrange
 multipliers $\{r_{i},\trho_{i}\}_{i=1}^{N}$. Here $r_{i}$ is a non-negative
number, and $\trho_{i}$ is a density operator. In general,
the optimal value of the dual problem differs from that of the
primal one. However, when the primal problem is convex and the
strictly feasible point of the parameters is located in the relative
interior of the problem domain, the optimal value of the primal and
dual problem coincides due to the sufficient condition of the strong
duality, called Slater's constraint qualification. Because our
optimization problem satisfies these conditions, the optimal duality
gap disappears \cite{ref:elda2}. By this property, the optimal
parameters of both the problems satisfy the  complementary slackness
$r_{i} \tr [\trho_{i} M_{i}]=0$($\forall i$). With this
condition, the constraints of primal problem and dual problem are
the necessary and sufficient conditions to optimality of two
problems. In optimization problem\cite{ref:boyd1,ref:berk1}, the optimality condition is
Karus-Khun-Tucker (KKT) condition. And the problem of the approach to
an optimal solution using these conditions is called LCP\cite{ref:bae1}. Using this way,
we obtain the optimal POVM and the set of complementary states, which provides optimal values to two problems.\\
\indent When $\{M_{i}\}_{i=1}^{N}$ is an optimal POVM with null operators $M_{j}=0(\forall j\notin\chi)$, $\{M_{i}\}_{i\in\chi}$ is a POVM which produces $p_{\rm guess}^{\chi}\equiv\max_{\{M_{i}'\}_{i\in\chi}}\sum_{i\in\chi}q_{i}\tr[\rho_{i}M_{i}']$ where $\chi$ is a subset of $\{1,2,\cdots,N\}$. And if we can optimize $\{q_{i},\rho_{i}\}_{i=1}^{N}$ through MD with less than $N$ quantum states, we have
\begin{equation}
p_{\g}=\max_{|\chi|=N-1}p_{\g}^{\chi}.\label{eq:pgchi}
\end{equation}
However if $\{q_{i},\rho_{i}\}_{i=1}^{N}$ is not optimized with MD
of less than $N$ quantum states, Eq.~(\ref{eq:pgchi}) cannot be used. At this
time we have
\begin{equation}
p_{\g}>\max_{1\leq i \leq N}q_{i}.
\end{equation}
It is because if $p_{\g}=q_{k}$, $\{M_{k}=I\}$
is an optimal POVM. Therefore we should know the condition to
classify two cases and the optimization method for MD with $N$
quantum states. In this paper $q_{1}$ denotes the largest prior
probability, which means $\max_{i}q_{i}=q_{1}$. This assumption does
not violate generality of our problem.\\
\indent For qubit states $(d=2)$, POVM elements and qubit states can
be represented by non-negative numbers $p_{i}$ and Bloch vectors
$\u_{i},\v_{i},\w_{i}$:
\begin{equation}\label{eq:georep}
M_{i}=p_{i}(I+{\u}_{i}\cdot\vec{\sigma}),
\rho_{i}=\frac{1}{2}(I+{\v}_{i}\cdot \vec{\sigma}),
\trho_{i}=\frac{1}{2}(I+{\w}_{i}\cdot \vec{\sigma}).
\end{equation}
When the representation is applied to KKT condition, the constraint
to the primal problem(or POVM constraint) is transformed to
\begin{equation}\label{eq:pcond}
p_{i}\geq 0~~ \forall i,~\sum_{i=1}^{N}p_{i}=1,
~\mathrm{and}~
~\sum_{i=1}^{N} p_{i}{\u}_{i}=\0,
\end{equation}
and the
constraint to the dual problem is expressed to
\begin{equation}\label{eq:dcond}
\tr[K]=q_{i}+r_{i}~~\forall i~~\mathrm{and}~~ \tr[K\cdot\vec{\sigma}]=q_{i}{\v}_{i}+r_{i}{\w}_{i}~~\forall i.
\end{equation}
And the complementary slackness
condition is represented as
\begin{equation}\label{eq:ccond}
p_{i}r_{i}(1+{\u}_{i}\cdot{\w}_{i})=0~~\forall i.
\end{equation}
\indent When every optimal POVM element is nonzero and its guessing
probability is greater than the largest prior probability
$q_{1}$, by Eqs.~\eqref{eq:georep},\eqref{eq:pcond},\eqref{eq:dcond}, we have
$p_{i},r_{i}>0(\forall i)$. Then the complementary slackness
condition is written as
$\|{\u}_{i}\|_{2}=1,{\w}_{i}=-{\u}_{i}(\forall i)$. We should note
that even in case that there exists such optimal POVM, there may
exist another optimal POVM with null operators. The KKT condition
\eqref{eq:pcond},\eqref{eq:dcond},\eqref{eq:ccond} can be expressed
using $p_{i},r_{i}>0(\forall i)$
\begin{eqnarray}\label{eq:geoc}
{\rm (i)}&&~~
p_{i},r_{i}>0~\forall i,~~\mbox{$\sum_{i=1}^{N}p_{i}=1$},~~\mbox{$\sum_{i=1}^{N}p_{i}\u_{i}=\0$}, \nonumber\\
{\rm (ii)}&&~~r_{i}-r_{1}=e_{i}~\forall i,~~r_{i}\u_{i}-r_{1}\u_{1}=\vec{s}_{i}~\forall i,\nonumber\\
{\rm (iii)}&&~~\|{\u}_{i}\|_{2} =1~\forall i,~~\w_{i}=-\u_{i}~\forall i,
\end{eqnarray}
where
\begin{equation}
e_{i}=q_{1}-q_{i}~~\mbox{and}~~\vec{s}_{i}=q_{i}\v_{i}-q_{1}\v_{1}.
\end{equation}
Reversely when $\{p_{i},{\u}_{i};r_{i},{\w}_{i}\}_{i=1}^{N}$ fulfills the above conditions,
$\{M_{i}\}_{i=1}^{N}$ and $\{r_{i},\trho_{i}\}_{i=1}^{N}$ corresponding to Eq.~\eqref{eq:georep} are an optimal POVM with $N$ nonzero elements and a set of complementary states, which provides the same optimal value $p_{\g}$ greater than the largest prior  probability $q_{1}$. Therefore Eq.~\eqref{eq:geoc} is a necessary and sufficient condition for every optimal POVM to have $N$ nonzero elements.\\
\indent Now to consider the mimimum error discrimination of four mixed qubit states with arbitrary prior probabilities, we explain geometric structure of $N$ qubit states. 

\subsection{Geometric Structure of $N$ qubit states}
In a set $\mathcal{C}$, we call the set of affine combinations of all
points as the {\em affine hull} of $\mathcal{C}$.
\begin{eqnarray}
\begin{array}{l}
\aff\,\mathcal{C}=\{\sum_{i}x_{i}c_{i}\,|\,c_{i}\in \mathcal{C}\, \forall i ,\, x_{i}\in\mathbb{R}\,\forall i ,\,\sum_{i}x_{i}=1\}.
\end{array}
\end{eqnarray}
And the dimension of $\aff\,\mathcal{C}$ is called as the {\em affine dimension} of $\mathcal{C}$. The geometric form of $\{q_{i},\rho_{i}\}_{i=1}^{N}$ can be classified by affine dimension of
$\{\vec{s}_{i}\}_{i=1}^{N}$. We denote the dimension as $D$. Since Bloch vectors lie in
three dimensional real space, we have $0\leq D \leq 3$. \\
\indent In $\mathcal{C}$, the set of convex combinations of every point is called as the {\em convex hull} of $\mathcal{C}$.
\begin{eqnarray}
\begin{array}{l}
\conv\,\mathcal{C}=\{\sum_{i}x_{i}c_{i}\,|\, c_{i}\in \mathcal{C}\,\forall i,\, x_{i}\geq 0\, \forall i ,\,\sum_{i}x_{i}=1\}.
\end{array}
\end{eqnarray}
$\conv\{\vec{s}_{i}\}_{i=1}^{N}$ is a point in $D=0$, a line
segment in $D=1$, a polygon in $D=2$, and a polyhedron in
$D=3$, respectively.\\ 
\indent When $\{p_{i},{\u}_{i};r_{i},{\w}_{i}\}_{i=1}^{N}$ satisfies
Eq.~(\ref{eq:geoc}), by the second condition of (ii), the affine
dimension of $\{r_{i}{\u}_{i}\}_{i=1}^{N}$ is $D$, and, by (i), the
relative interior of $\conv\{r_{i}{\u}_{i}\}_{i=1}^{N}$ contains $\0$ which is the
origin of the Bloch ball. If $D<N-1$, by Carath\'{e}odory's
theorem\cite{ref:berk1}, there is $\{\tilde{p}_{i}\}_{i=1}^{N}$ with $D+1$ or fewer nonzero
elements such that $\{\tilde{p}_{i},\u_{i}\}_{i=1}^{N}$ satisfies
$\tilde{p}_{i}\geq0(\forall i)$, $\sum_{i=1}^{N}\tilde{p}_{i}=1$,
and $\sum_{i=1}^{N}\tilde{p}_{i}r_{i}\u_{i}=\0$. Then
$\{p_{i}'=\tilde{p}_{i}r_{i}/\sum_{j=1}^{N}\tilde{p}_{j}r_{j},\u_{i}\}_{i=1}^{N}$
satisfies Eq.~(\ref{eq:pcond}). This means that there exists an
optimal POVM with $D+1$ or fewer nonzero elements because
$\{p_{i}',\u_{i};r_{i},{\w}_{i}\}_{i=1}^{N}$ satisfies KKT condition
\eqref{eq:pcond},\eqref{eq:dcond},\eqref{eq:ccond}. Therefore, in the case of 
$D=0$, by $N\geq 2$, we have $p_{\g}=q_{1}$ and in the case of $N>4$, by
$D\leq 3$, $\{q_{i},\rho_{i}\}_{i=1}^{N}$ is optimized with MD of
less than $N$ quantum states\cite{ref:davi1,ref:hunt1}. Since
$D>N-1$ is impossible, if $D\neq N-1$, we can obtain the guessing
probability through MD with less than $N$ quantum states.
\subsection{Geometric optimality condition for $N$ qubit states}
Let us consider the MD of $\{q_{i},\rho_{i}\}_{i=1}^{N}$ with $D=N-1$.
When $\{p_{i},{\u}_{i};r_{i},{\w}_{i}\}_{i=1}^{N}$
fulfills Eq.~\eqref{eq:geoc}, the polytope $\conv\{\vec{s}_{i}\}_{i=1}^{N}$ becomes a $D$-dimensional simplex, which is a
line segment in $N=2$ or a triangle in $N=3$, or a tetrahedron
in $N=4$. Then the second condition of (ii) of Eq.~\eqref{eq:geoc}
means that $\conv\{r_{i}{\u}_{i}\}_{i=1}^{N}$ is congruent to
$\conv\{\vec{s}_{i}\}_{i=1}^{N}$ and can be overlapped by parallel
transport $\co$. This implies the following facts. Firstly, the distance from
$r_{1}\u_{1}$ to $r_{i}\u_{i}$ is the same as $l_{i}$, which is the
distance between $q_{1}\v_{1}$ and $q_{i}\v_{i}$, that is,
$l_{i}=\|\vec{s}_{i}\|_{2}$. Secondly, the \emph{relative interior} of $\conv\{\vec{s}_{i}\}_{i=1}^{N}$, $\Omega$, contains $\co$ because the
relative interior of $\conv\{r_{i}\u_{i}\}_{i=1}^{N}$ contains $\0$. Note that $\Omega$ has infinite numbers of elements.
Thirdly,
\begin{equation}
r_{i}=\|\vec{s}_{i}-\co\|_{2},\
\u_{i}=\frac{\vec{s}_{i}-\co}{\|\vec{s}_{i}-\co\|_{2}}=-\w_{i}.
\end{equation}
Since $\conv\{\vec{s}_{i}\}_{i=1}^{N}$ is
a simplex, for any $\c\in\Omega$, there exists
a unique $\{t_{i}(\c)\}_{i=1}^{N}$ which satisfies the following relations.
\begin{equation}\label{eq:tcond}
t_{i}(\c)>0~\forall i,~~\sum_{i=1}^{N}t_{i}(\c)=1,~~\sum_{i=1}^{N}t_{i}(\c)\vec{s}_{i}=\c,
\end{equation}
which implies that 
\begin{equation}
p_{i}=\frac{t_{i}(\co)\|\vec{s}_{i}-\co\|_{2}}{\sum_{j=1}^{N}t_{j}(\co)\|\vec{s}_{j}-\co\|_{2}}.
\end{equation}
In fact, for any $\c\in\Omega$, $\{p_{i},\u_{i};r_{i},\w_{i}\}_{i=1}^{N}$, defined in the
following way, satisfies (i) and (iii) of Eq.~\eqref{eq:geoc}, and the second condition of (ii).
\begin{equation}\label{eq:purw}
\begin{array}{ll}
p_{i}=\frac{t_{i}(\c)\|\vec{s}_{i}-\c\|_{2}}{\sum_{j=1}^{N}t_{j}(\c)\|\vec{s}_{j}-\c\|_{2}},~~&
\u_{i}=\frac{\vec{s}_{i}-\c}{\|\vec{s}_{i}-\c\|_{2}},\\
r_{i}=\|\vec{s}_{i}-\c\|_{2},&
\w_{i}=\frac{\c-\vec{s}_{i}}{\|\vec{s}_{i}-\c\|_{2}}.
\end{array}
\end{equation}
Therefore $\{p_{i},\u_{i};r_{i},\w_{i}\}_{i=1}^{N}$ satisfying 
Eq.~\eqref{eq:geoc} except the first
condition of (ii) are always innumerable.
$\co$ is an optimal $\c\in\Omega$ which satisfies
up to the first condition of (ii) of Eq.~\eqref{eq:geoc}.\\ 
\indent The meaning of the first condition of (ii) can be understood as
$\co\in\bigcap_{i=2}^{N}\C_{i}$ 
because $r_{i}$ is $\|\vec{s}_{i}-\c\|_{2}$ in Eq.~\eqref{eq:purw}. Here $\C_{i}$ is the hyperboloid made
of points such that difference between the distance $\vec{s}_{i}$
and $\0$ is $e_{i}$. That is,
\begin{equation}
\C_{i}=\{\c\in\mathbb{R}^{3}:
\|\c-\vec{s}_{i}\|_{2}-\|\c\|_{2}=e_{i}
\}~~\forall i\in\I,
\end{equation}
where
\begin{equation}
\I=\{i\in\mathbb{Z}:2\leq i \leq N\}.
\end{equation}
This implies that if $\c\in(\bigcap_{i\in\I}{\C}_{i})\cap\Omega$, $\{p_{i},\u_{i};r_{i},\w_{i}\}_{i=1}^{N}$ of Eq.~\eqref{eq:purw} satisfies every condition in Eq.~\eqref{eq:geoc}, that is, $\c=\co$.
Furthermore if $(\bigcap_{i\in\I}{\C}_{i})\cap\Omega$ is
empty, there does not exist $\{p_{i},{\u}_{i};r_{i},{\w}_{i}\}_{i=1}^{N}$ satisfying Eq.~\eqref{eq:geoc}. In
this case, $\{q_{i},\rho_{i}\}_{i=1}^{N}$ is optimized with MD of
less than $N$ quantum states. Therefore $\{q_{i},\rho_{i}\}_{i=1}^{N}$ is not optimized with MD of less than
$N$ quantum states if and only if
\begin{equation}\label{eq:nozec}
\begin{array}{c}
D=N-1~~{\rm and}~~(\bigcap_{i\in\I}{\C}_{i})\cap\Omega\neq\emptyset.
\end{array}
\end{equation}
In addition, using $\Theta_{i}$ which is the angle between two segments
$\conv\{\0,\cw\}$ and $\conv\{\0,\vec{s}_{i}\}$, we can obtain the following relation by
hyperbolic equation.
\begin{equation}\label{eq:radius}
\|\cw\|_{2}=\frac{l_{i}^{2}-e_{i}^{2}}{2(l_{i}\cos\Theta_{i}+e_{i})}\quad\forall
i\in\I.
\end{equation}
Here $\cw$ represents an element of $(\bigcap_{i\in\I}{\C}_{i})\cap\Lambda$ and is a potential candidate of $\co$, where $\Lambda$ is the relative interior of convex cone
whose apex(base) is $\0$($\conv\{\vec{s}_{i}\}_{i\in\I}$).
Since $\|\cw\|_{2}$ is strictly increasing function of $\Theta_{i}$, $\cw$ and $\{\Theta_{i}\}_{i\in\I}$ are  unique. By $\Omega\subset\Lambda$, if $\cw\in\Omega$, we obtain  $\co=\cw$. This means that $\co$ is unique. Therefore if Eq.~\eqref{eq:nozec} holds, the optimal POVM is unique, and the guessing probability $p_{\g}$ is $q_{1}+\|\cw\|_{2}$ by $r_{1}=\|\cw\|_{2}$. The following theorem summarizes our results obtained in this section.
\begin{theorem}\label{the:geo}
When $D=N-1$, $N$ qubit states $\{q_{i},\rho_{i}\}_{i=1}^{N}$ is not optimized with MD
of less than $N$ quantum states if and only if $\exists\cw\in\Omega$. Then the minimum-error measurement
is unique, and the guessing probability is $q_{1}+\|\cw\|_{2}$.
The optimal POVM and the set of complementary states are obtained by
substituting Eq.~\eqref{eq:purw} of $\c=\cw$ into Eq.~\eqref{eq:georep}.
\end{theorem}
Note that the semidefinite programming of MD or the
derivation of Theorem \ref{the:geo} is independent of the normalization of prior probabilities. In other words, Theorem \ref{the:geo} can be applied
even to the case of un-normalized prior probabilities.

\section{Optimal discrimination of four qubit states}
\label{sec:unequal4}

Now, we will explain $\exists\cw\in\Omega$ in terms of an analytic form. First of all, by the
geometric condition $D=N-1$, we classify our
problem into the cases of $N=2,3$, and $4$. In $N=2$,
$\conv\{\vec{s}_{i}\}_{i=1}^{N}$ is a line segment. In $N=3$, it
is a triangle. And in $N=4$, it becomes a tetrahedron.
 To avoid the repetition of
representation at $N=3$ or $4$, we use different indices $x,y,z$. That is,
when $N=3$, we use only
$x,y\in\I$. However, when $N=4$, we use $x,y,z\in\I$. The process to
check the existence of $\co$ can be understood in three steps. At
first, we confirm the existence of $\{\C_{i}\cap\Lambda\}_{i\in\I}$.
Secondly, we check the existence of the intersection point $\cw$.
Thirdly, we verify $\cw\in\Omega$. Then we can see that $\cw$ becomes $\co$.
\subsection{Necessary and sufficient condition for no null measurement operator}
\begin{figure}[!t]
\centerline{\includegraphics[scale=0.32]{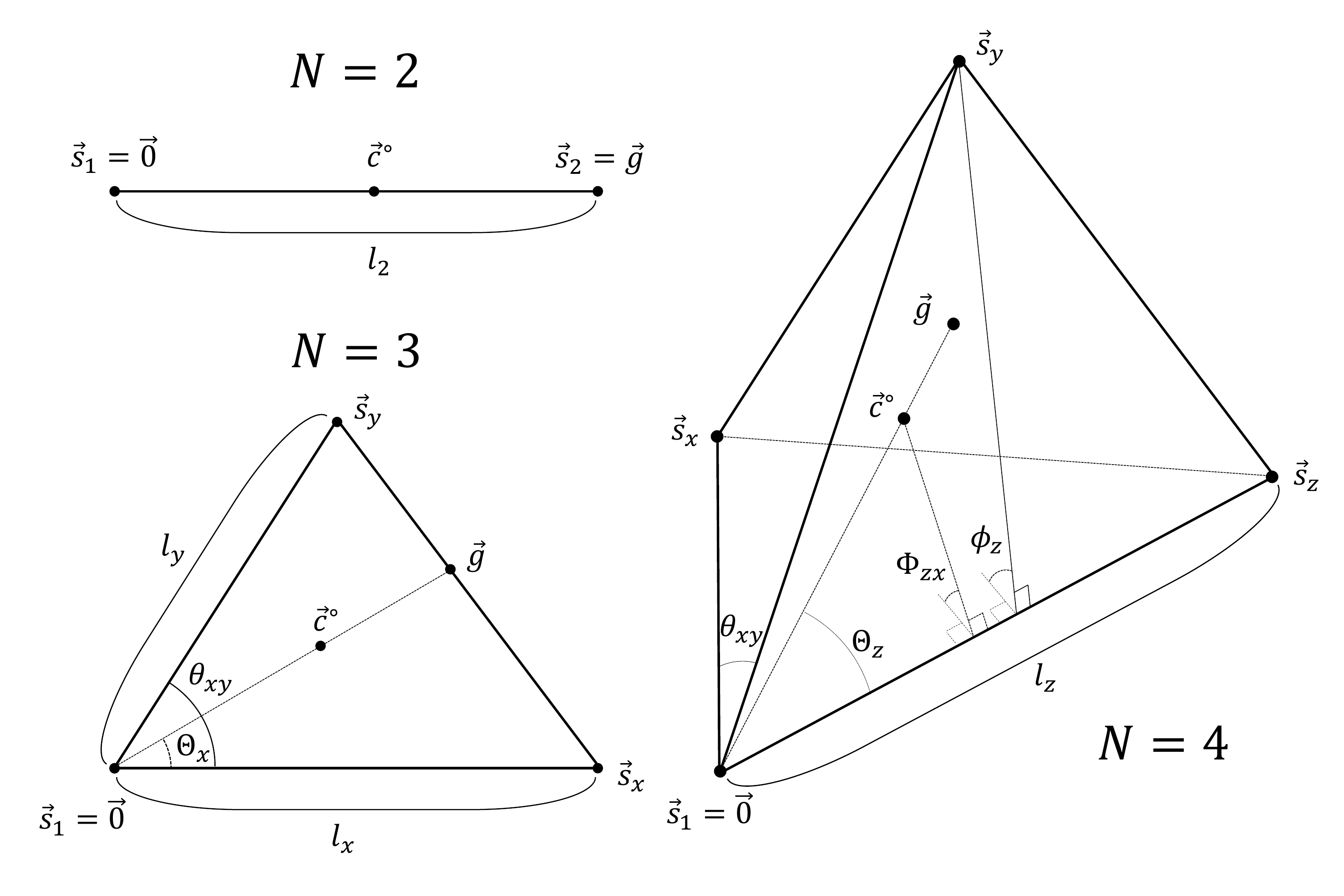}}
\caption{$\cw\in\Omega$ when $N=2,3,4$. The geometric condition
$D=N-1$ means that when $N=2$, $\conv\{\vec{s}_{i}\}_{i=1}^{N}$ is a
line segment, and when $N=3$, it is a triangle, and when $N=4$, it
becomes a tetrahedron. The other geometric condition
$(\bigcap_{i\in\I}\C_{i})\cap\Omega\neq\emptyset$ implies that
$\bigcap_{i\in\I}(\C_{i}\cap\Lambda)$ is a nonempty set and its element
$\cw$ should be included in $\Omega$. $\cw\in\Omega$ can be
expressed as $\|\cw\|_{2}<\|\vg\|_{2}$. If the condition is
satisfied, then $\cw$ becomes $\co$.} \label{fig:tetra}
\end{figure}

\indent The first step can be represented, by Eq.~\eqref{eq:radius}, as
the following analytic form:
\begin{equation}\label{eq:excur}
l_{i}>e_{i}\quad\forall i\in\I.
\end{equation}
In the case of $N=2$, $\bigcap_{i\in\I}(\C_{i}\cap\Lambda)$ becomes ${\C}_{2}\cap\Lambda$, and the element of the set is located in ${\rm \Omega}$.
Therefore $\exists\cw\in\Omega$ can be completely represented by Eq.~\eqref{eq:excur}. However it is not the cases when $N=3$ or $4$ since even in the case where $\{{\C}_{i}\cap\Lambda\}_{i\in\I}$ may exist, $\cw$ may not exist and even when $\cw$ exist, it may not be the element of $\Omega$.\\
\indent To express the second step in an analytic form, we have to
find an analytic form of $\{\Theta_{i}\}_{i\in\I}$, satisfying Eq.~\eqref{eq:radius}.
If $N=2$,   $\Theta_{2}=0$ holds trivially. However, if $N=3$ or $4$, we
substitute the following relations into Eq.~\eqref{eq:radius}.
\begin{equation}
\begin{array}{rcl}
\cos\Theta_{y}&=&\cos\Theta_{x}\cos\theta_{xy}
+\sin\Theta_{x}\sin\theta_{xy}\cos\Phi_{xy},\\[2mm]
\sin^{2}\phi_{z}&=&\cos^{2}\Phi_{zx}+\cos^{2}\Phi_{zy}\\
&&-2\cos\phi_{z}\cos\Phi_{zx}\cos\Phi_{zy},
\end{array}
\end{equation}
where $\theta_{xy}$ is the internal angle between two line segments
$\conv\{\vec{s}_{1},\vec{s}_{x}\}$ and
$\conv\{\vec{s}_{1},\vec{s}_{y}\}$, $\phi_{z}$ is
the internal angle between two triangles
$\conv\{\vec{s}_{1},\vec{s}_{z},\vec{s}_{x}\}$ and
$\conv\{\vec{s}_{1},\vec{s}_{z},\vec{s}_{y}\}$, 
and $\Phi_{xy}$ is
the internal angle between two triangles
$\conv\{\vec{s}_{1},\vec{s}_{x},\cw\}$ and
 $\conv\{\vec{s}_{1},\vec{s}_{x},\vec{s}_{y}\}$. When $N=3$, we get
$\Theta_{i}=\alpha_{i}(\forall i \in\I)$:
\begin{equation}\label{eq:alpha}
\begin{array}{c}
\alpha_{x}=\arccos\left(\frac{-X_{xy}Y_{xy}+\sqrt{1+X_{xy}^{2}-Y_{xy}^{2}}}{1+X_{xy}^{2}}\right),
\end{array}
\end{equation}
where
\begin{equation}\label{eq:xy}
\begin{array}{rcl}
X_{xy}&=&\frac{l_{x}(l_{y}^{2}-e_{y}^{2})-l_{y}(l_{x}^{2}-e_{x}^{2})\cos \theta_{xy}}{l_{y}(l_{x}^{2}-e_{x}^{2})\sin \theta_{xy}},\\
Y_{xy}&=&\frac{e_{x}(l_{y}^{2}-e_{y}^{2})-e_{y}(l_{x}^{2}-e_{x}^{2})}{l_{y}(l_{x}^{2}-e_{x}^{2})\sin\theta_{xy}}.
\end{array}
\end{equation}
In the case of $N=4$, we have $\Theta_{i}=\beta_{i}(\forall i\in\I)$:
\begin{eqnarray}\label{eq:beta}
\begin{array}{c}
\beta_{z}=\arccos\left(\frac{-\bar{Z}_{z}+\sqrt{\bar{Z}_{z}^{2}-\big(\bar{X}_{z}+\sin^{2}\phi_{z}\big)\big(\bar{Y}_{z}-\sin^{2}\phi_{z}\big)}}{\bar{X}_{z}+\sin^{2}\phi_{z}}\right),
\end{array}
\end{eqnarray}
where
\begin{equation}
\begin{array}{rcl}
\bar{X}_{z}&=&X_{zx}^{2}+X_{zy}^{2}-2X_{zx}X_{zy}\cos \phi_{z},\\
\bar{Y}_{z}&=&Y_{zx}^{2}+Y_{zy}^{2}-2Y_{zx}Y_{zy}\cos \phi_{z},\\
\bar{Z}_{z}&=&X_{zx}Y_{zx}+X_{zy}Y_{zy}\\
&&-(X_{zx}Y_{zy}+Y_{zx}X_{zy})\cos
\phi_{z}.
\end{array}
\end{equation}
Then, $\Phi_{xy}$ is given as the following $\gamma_{xy}$:
\begin{eqnarray}\label{eq:gamma}
\gamma_{xy}&=&\arccos\left(\frac{X_{xy}\cos\beta_{x}+Y_{xy}}{\sin\beta_{x}}\right).
\end{eqnarray}
If Eq.~\eqref{eq:excur} is fulfilled, the value inside the root of
$\alpha_{x}$ is always positive however that of $\beta_{z}$ is not
always positive. Therefore unlike $N=3$ we should add the following
condition in the case of $N=4$:
\begin{eqnarray}\label{eq:addcond}
\bar{Z}_{i}^{2}\geq\big(\bar{X}_{i}+\sin^{2}\phi_{i}\big)\big(\bar{Y}_{i}-\sin^{2}\phi_{i}\big)~~\forall i\in\I.
\end{eqnarray}
In fact points corresponding to $\{\alpha_{i}\}_{i\in\I}$ at $N=3$
and points corresponding to $\{\beta_{i}\}_{i\in\I}$ at $N=4$ are
located in $(\bigcap_{i\in\I}\C_{i})\cap\aff\{\vec{s}_{i}\}_{i=1}^{N}$.
By $\Lambda\subset\aff\{\vec{s}_{i}\}_{i=1}^{N}$, this implies that
if one of them is included in $\Lambda$, the point is $\cw$, but if
one of them is not included in $\Lambda$, $\cw$ does not exist.
 Therefore, for $\alpha_{x}$ and $\alpha_{y}$ at
$N=3$, we have
\begin{eqnarray}\label{eq:alphacond}
\alpha_{x},\alpha_{y}<\theta_{xy}~~{\rm and}~~\alpha_{x}+\alpha_{y}<\pi.
\end{eqnarray}
If we consider $\{(\gamma_{zx},\gamma_{zy})\}_{z}$ at $N=4$, we find
\begin{eqnarray}\label{eq:gammacond}
\gamma_{zx},\gamma_{zy}<\phi_{z}~~{\rm and}~~\gamma_{zx}+\gamma_{zy}<\pi\quad
\forall z\in\I.
\end{eqnarray}
\indent Final stage can be represented as the relation between
$\|\cw\|_{2}$ and $\|\vg\|_{2}$. $\vg$ denotes the intersection point of
two sets $\aff\{\vec{s}_{1},\cw\}$ and  $\conv\{\vec{s}\}_{i\in\I}$.
The inequality of $\|\cw\|_{2}<\|\vg\|_{2}$ implies $\cw\in\Omega$,
and the inequality $\|\cw\|_{2}\geq\|\vg\|_{2}$ means
$\cw\notin\Omega$. Since $\conv\{\vec{s}_{i}\}_{i=1}^{N}$ is a simplex,
$\conv\{\vec{s}_{i}\}_{i\in\I}$ becomes a simplex. It means that there
exists a unique $\{\bar{t}_{i}\}_{i\in\I}$ satisfying the following
relations
\begin{equation}
\bar{t}_{i}>0~~\forall i\in\I,~\sum_{i\in\I}\bar{t}_{i}=1,~\sum_{i\in\I}\bar{t}_{i}\vec{s}_{i}=\vg.
\end{equation}
Then $\{\bar{t}_{i}\}_{i\in\I}$ and $\|\vg\|_{2}$ are expressed as follows:
\begin{equation}\label{eq:tvg}
\begin{array}{rcl}
N=2:~&&\bar{t}_{2}=1,~~\|\vg\|_{2}=l_{2},\\[3mm]
N=3:~&&\bar{t}_{x}=\frac{l_{y}\sin\alpha_{y}}{l_{x}\sin\alpha_{x}+l_{y}\sin\alpha_{y}},\nonumber\\
~&&\|\vg\|_{2}=\frac{l_{x}l_{y}\sin\theta_{xy}}{l_{x}\sin\alpha_{x}+l_{y}\sin\alpha_{y}},\\[3mm]
N=4:~&&\bar{t}_{z}=\frac{\triangle_{z}\Gamma_{z}}{\triangle_{x}\Gamma_{x}+\triangle_{y}\Gamma_{y}+\triangle_{z}\Gamma_{z}},\\
~&&
\|\vg\|_{2}=\frac{3\triangle_{\rm
tetra}}{\triangle_{x}\Gamma_{x}+\triangle_{y}\Gamma_{y}+\triangle_{z}\Gamma_{z}},
\end{array}
\end{equation}
where
\begin{eqnarray}
\Gamma_{z} = \sin\beta_{x}\sin\gamma_{xy}.
\end{eqnarray}
Here $\triangle_{z}$ is the area of triangle $\conv\{\vec{s}_{1},\vec{s}_{x},\vec{s}_{y}\}$ and $\triangle_{\rm tetra}$ is the volume of tetrahedron $\conv\{\vec{s}_{i}\}_{i=1}^{4}$. Therefore if $\exists\cw\in\Omega$ holds, $\co$ and $\{t_{i}(\co)\}_{i=1}^{N}$ can be expressed as follows;
\begin{equation}\label{eq:cot}
\begin{array}{rcl}
\co&=&\frac{\|\cw\|_{2}}{\|\vg\|_{2}}\cdot \sum_{i\in\I}\bar{t}_{i}\vec{s}_{i},\\
t_{1}(\co)&=&1-\frac{\|\cw\|_{2}}{\|\vg\|_{2}},\\
t_{i}(\co)&=&\frac{\|\cw\|_{2}}{\|\vg\|_{2}}\cdot \bar{t}_{i}~~\forall i\in\I.
\end{array}
\end{equation}
Figure \ref{fig:tetra} shows $\cw$ when $D=N-1$ and $\exists\cw\in\Omega$.
Now, we use $m$ to denote the maximal number of null operators which can exist in optimal measurement. Table \ref{tab:anal} in appendix provides the analyric form of necessary and sufficient condition for the case of $m=0$ in $D=N-1$ of $N$ qubit states.
Briefly, $l_{2}>e_{2}$ is the condition of two qubit states in the case of $N=2$ and $D=1$.
{\sf C1}({\sf C2}) is the condition for  $N=3$ and $D=2$($N=4$ and $D=3$) in three(four) qubit states.

The following theorem provides the complete analysis to the mimimum error discrimination of four mixed qubit states with arbitrary prior probabilities.

\begin{theorem}\label{the:4null}
When arbitrary four qubit states $\{q_{i},\rho_{i}\}_{i=1}^{4}$ with $D=3$ are given, every optimal POVM has four nonzero elements if and only if
\begin{equation}\label{eq:nsc4}
\begin{array}{l}
l_{2}>e_{2},\,l_{3}>e_{3},\,l_{4}>e_{4},\\[2mm]
\bar{Z}_{2}^{2}\geq\big(\bar{X}_{2}+\sin^{2}\phi_{2}\big)\big(\bar{Y}_{2}-\sin^{2}\phi_{2}\big),\\
\bar{Z}_{3}^{2}\geq\big(\bar{X}_{3}+\sin^{2}\phi_{3}\big)\big(\bar{Y}_{3}-\sin^{2}\phi_{3}\big),\\
\bar{Z}_{4}^{2}\geq\big(\bar{X}_{4}+\sin^{2}\phi_{4}\big)\big(\bar{Y}_{4}-\sin^{2}\phi_{4}\big),\\[2mm]
\gamma_{23},\gamma_{24}<\phi_{2}~~{\rm and}~~\gamma_{23}+\gamma_{24}<\pi,\\
\gamma_{34},\gamma_{32}<\phi_{3}~~{\rm and}~~\gamma_{34}+\gamma_{32}<\pi,\\
\gamma_{42},\gamma_{43}<\phi_{4}~~{\rm and}~~\gamma_{42}+\gamma_{43}<\pi,\\[2mm]
\frac{l_{2}^{2}-e_{2}^{2}}{2(l_{2}\cos\beta_{2}+e_{2})}<
\frac{3\triangle_{\rm tetra}}{\triangle_{2}\Gamma_{2}+\triangle_{3}\Gamma_{3}+\triangle_{4}\Gamma_{4}}.
\end{array}
\end{equation}
Then $\{p_{i},\u_{i}\}_{i=1}^{4}$ of the unique optimal POVM and guessing probability are given by 
\begin{eqnarray*}\label{eq:4optme}
\begin{array}{rcl}
p_{1}&=&\frac{6\triangle_{\rm tetra}(l_{2}\cos\beta_{2}+e_{2})-(l_{2}^{2}-e_{2}^{2})\sum_{j=2}^{4}\triangle_{j}\Gamma_{j}}
{2(l_{2}\cos\beta_{2}+e_{2})(3\triangle_{\rm tetra}+\sum_{k=2}^{4}e_{k}\triangle_{k}\Gamma_{k})},\\[3mm]
\u_{1}&=&-\frac{\triangle_{2}\Gamma_{2}\cdot\vec{s}_{2}+\triangle_{3}\Gamma_{3}\cdot\vec{s}_{3}+\triangle_{4}\Gamma_{4}\cdot\vec{s}_{4}}{3\triangle_{\rm tetra}},\\[3mm]
p_{i}&=&\frac{\triangle_{i}\Gamma_{i}(l_{i}^{2}+e_{i}^{2}+2l_{i}e_{i}\cos\beta_{i})}
{2(l_{i}\cos\beta_{i}+e_{i})(3\triangle_{\rm tetra}+\sum_{k=2}^{4}e_{j}\triangle_{j}\Gamma_{j})},~\forall i\neq 1,\\[3mm]
\u_{i}&=&\frac{6\triangle_{\rm tetra}(l_{i}\cos\beta_{i}+e_{i})\vec{s}_{i}-(l_{i}^{2}-e_{i}^{2})\sum_{j=2}^{4}\triangle_{j}\Gamma_{j}\cdot\vec{s}_{j}}
{3\triangle_{\rm tetra}(l_{i}^{2}+e_{i}^{2}+2l_{i}e_{i}\cos\beta_{i})},
\end{array}
\end{eqnarray*}
and
\begin{equation}\label{eq:4guess}
p_{\g}=q_{1}+\frac{l_{2}^{2}-e_{2}^{2}}{2(l_{2}\cos\beta_{2}+e_{2})}.
\end{equation}
\end{theorem}

\begin{figure}[!t]
\centerline{\includegraphics*[bb=30 30 380 380,scale=0.65]{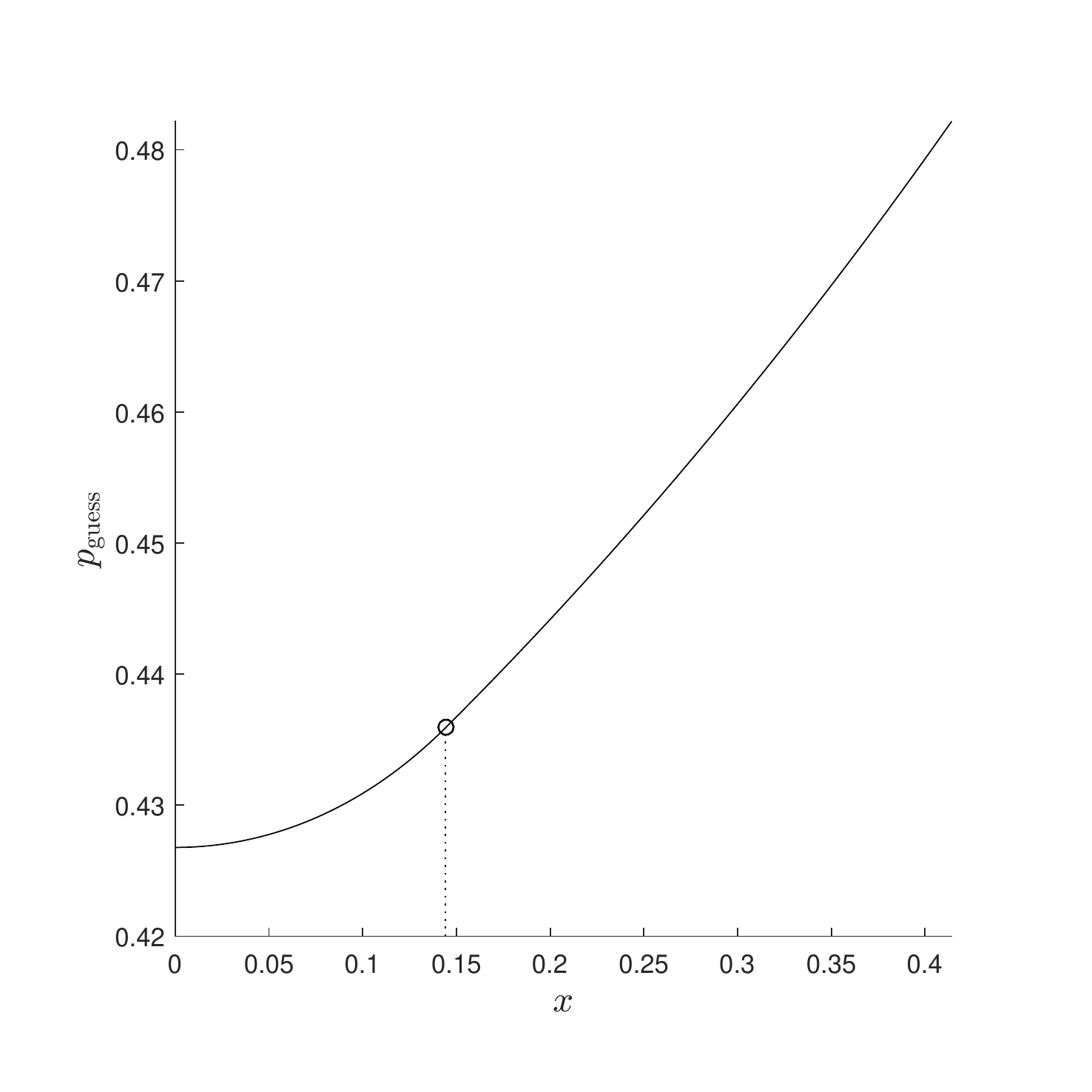}}
\caption{Guessing probability of non-equiprobable non-symmetric four qubit states. If $0\leq h<h^{\star}\approx0.1440$, we get four elements optimal POVM. However, if $h^{\star}\leq h\leq \sqrt{2}-1\approx0.4143$, we obtain three elements optimal POVM.}\label{fig:xvspg4}
\end{figure}

\indent Here let us consider an example of minimum error discrimination to mixed four qubit states with non-equal prior probabilities. To understand the behavior of minimum error discrimination according to prior probabilities,
we consider non-fixed prior probabilities. The mixed four qubit states and the prior probabilities are as follows:  
\begin{equation}
\begin{array}{ll}
q_{1}=\frac{1+h}{4},&\v_{1}=\frac{1}{2}(1,0,-1),\\[2mm]
q_{2}=\frac{1}{4},&\v_{2}=\frac{1+h}{2}(-1,0,-1),\\[2mm]
q_{3}=\frac{1}{4},&\v_{3}=\frac{1-h}{2}(0,1,1),\\[2mm]
q_{4}=\frac{1-h}{4},&\v_{4}=\frac{1}{2}(0,-1,1),\\[2mm]
&0\leq h \leq \sqrt{2}-1\approx0.4142.
\end{array}
\end{equation}
When $h=0$, the prior probabilities are identical and $\{\v_{i}\}_{i=1}^{4}$ forms a symmetric tetrahedron of which circumcenter and circumradius are $\0$ and $\frac{1}{\sqrt{2}}$, respectively. Then the guessing probability becomes $\frac{1}{4}+\frac{1}{4\sqrt{2}}$. It is because if the purify $f_{i}(=\|\v_{i}\|_{2})$ of equiprobable four qubit states is $f$, the origin $\0$ is included in relative interior of $\conv\{v\}_{i=1}^{4}$ and the guessing probability becomes $\frac{1}{4}+\frac{1}{4}f$\cite{ref:bae3}.\\ 
\indent However, when $0<h\leq\sqrt{2}-1$, we have $q_{1}>q_{2}=q_{3}>q_{4}$ and $\{\v_{i}\}_{i=1}^{4}$ constructs a non-symmetric tetrahedron. In addition,
purity of four qubit states cannot be the same, because of $f_{2}>f_{1}=f_{4}>f_{3}$.
Therefore, without using the method proposed in this paper, guessing probability cannot be obtained. Once we use the result of section 3.2, we can obtain the guessing probability. 
\begin{widetext}
\begin{equation}
p_{\rm guess}=\left\{
\begin{array}{lll}
\frac{1}{4}+\frac{h}{4}+\frac{1-4h^{2}+2h^{3}+12h^{4}+4h^{5}-h^{6}+2h^{7}+2h^{8}}{4h(2-h-10h^{2}-2h^{3}+2h^{4}-h^{5}-2h^{6})+4(1-h^{2})\sqrt{2-10h^{2}+5h^{4}-h^{8}}}&{\rm if}&h< h^{\star},\\
\frac{1}{4}+\frac{h}{4}+\frac{9+18h+h^{2}-8h^{3}+13h^{4}+6h^{5}+h^{6}}{8h(7+10h-6h^{2}-2h^{3}-h^{4})+8(1+h)\sqrt{(1+2h)(9-h^{4})(5-2h+h^{2})}}&{\rm if}&h \geq h^{\star}.
\end{array}
\right.
\end{equation}
\end{widetext}
$h^{\star}(\approx0.1440)$ is the value which is the boundary between four elements optimal POVM and three elements optimal POVM. If $0\leq h<h^{\star}$, we get four elements optimal POVM. However, if $h^{\star}\leq h\leq \sqrt{2}-1$, we obtain three elements optimal POVM. Figure \ref{fig:xvspg4} shows $p_{\g}(h)$ intuitively.\\
\begin{table}[tt]
\caption{The necessary and sufficient condition that every optimal measurement has 
no null operator for $N$ qubit states in $D=N-1$. $\exists\cw\in\Omega$ has different analytic form for $N=2,3,4$. Note that $q_{1}$ is the largest prior probability.}\label{tab:anal}
\renewcommand{\arraystretch}{2}
\begin{tabular}{ccc}
&{\bf Geometric form}&{\bf Analytic form}\\
$N=2$&$\C_{2}\cap\Lambda\neq\emptyset$&$l_{2}>e_{2}$\\
({\sf C0}) & &  \\
\hline \\
\multirow{3}{*}{\begin{tabular}{c}$N=3$\\({\sf C1})\end{tabular}}
&$\C_{2}\cap\Lambda,\,\C_{3}\cap\Lambda\neq\emptyset$&$l_{2}>e_{2},\,l_{3}>e_{3}$\\
&$\exists\cw$ when $\C_{2}\cap\Lambda,\,\C_{3}\cap\Lambda\neq\emptyset$&$\alpha_{2},\alpha_{3}<\theta_{23}$ and $\alpha_{2}+\alpha_{3}<\pi$\\
&$\cw\in\Omega$ when $\exists\cw$&$
\frac{l_{2}^{2}-e_{2}^{2}}{2(l_{2}\cos\alpha_{2}+e_{2})}
<\frac{l_{2}l_{3}\sin\theta_{23}}{l_{2}\sin\alpha_{2}+l_{3}\sin\alpha_{3}}$\\
\hline \\
&$\C_{2}\cap\Lambda,\,\C_{3}\cap\Lambda,\,\C_{4}\cap\Lambda\neq\emptyset$&$l_{2}>e_{2},\,l_{3}>e_{3},\,l_{4}>e_{4}$\\
&\multirow{6}{*}{$\exists\cw$ when $\C_{2}\cap\Lambda,\,\C_{3}\cap\Lambda,\,\C_{4}\cap\Lambda\neq\emptyset$}
 &$\bar{Z}_{2}^{2}\geq\big(\bar{X}_{2}+\sin^{2}\phi_{2}\big)\big(\bar{Y}_{2}-\sin^{2}\phi_{2}\big)$\\
&&$\bar{Z}_{3}^{2}\geq\big(\bar{X}_{3}+\sin^{2}\phi_{3}\big)\big(\bar{Y}_{3}-\sin^{2}\phi_{3}\big)$\\
$N=4$&&$\bar{Z}_{4}^{2}\geq\big(\bar{X}_{4}+\sin^{2}\phi_{4}\big)\big(\bar{Y}_{4}-\sin^{2}\phi_{4}\big)$\\
({\sf C2})&&$\gamma_{23},\gamma_{24}<\phi_{2}~~{\rm and}~~\gamma_{23}+\gamma_{24}<\pi$\\
&&$\gamma_{34},\gamma_{32}<\phi_{3}~~{\rm and}~~\gamma_{34}+\gamma_{32}<\pi$\\
&&$\gamma_{42},\gamma_{43}<\phi_{4}~~{\rm and}~~\gamma_{42}+\gamma_{43}<\pi$\\
&$\cw\in\Omega$ when $\exists\cw$&$\frac{l_{2}^{2}-e_{2}^{2}}{2(l_{2}\cos\beta_{2}+e_{2})}<
\frac{3\triangle_{\rm
tetra}}{\triangle_{2}\Gamma_{2}+\triangle_{3}\Gamma_{3}+\triangle_{4}\Gamma_{4}}$\\
\hline \\
$e_{i},\vec{s}_{i},l_{i}$&
\multicolumn{2}{c}{$e_{i}=q_{1}-q_{i}$,~$\vec{s}_{i}=q_{i}\v_{i}-q_{1}\v_{1}$,~and~$l_{i}=\|\vec{s}_{i}\|_{2}$}\\
$x,y,z$&\multicolumn{2}{c}{$\{x,y\}=\{2,3\}$ when $N=3$,~~$\{x,y,z\}=\{2,3,4\}$ when $N=4$}\\
$\theta_{xy}$&
\multicolumn{2}{c}{
angle between line segments
$\conv\{\vec{s}_{1},\vec{s}_{x}\}$ and
$\conv\{\vec{s}_{1},\vec{s}_{y}\}$ when $N=3,4$}\\
$\phi_{z}$&
\multicolumn{2}{c}{
angle between triangles
$\conv\{\vec{s}_{1},\vec{s}_{z},\vec{s}_{x}\}$ and
$\conv\{\vec{s}_{1},\vec{s}_{z},\vec{s}_{y}\}$ when $N=4$}\\
$\triangle_{z},\triangle_{\rm tetra}$&\multicolumn{2}{c}{
area of triangle $\conv\{\vec{s}_{1},\vec{s}_{x},\vec{s}_{y}\}$,~
volume of tetrahedron $\conv\{\vec{s}_{i}\}_{i=1}^{4}$}\\
$X_{xy},Y_{xy}$&\multicolumn{2}{c}{
$\frac{l_{x}(l_{y}^{2}-e_{y}^{2})-l_{y}(l_{x}^{2}-e_{x}^{2})\cos \theta_{xy}}{l_{y}(l_{x}^{2}-e_{x}^{2})\sin \theta_{xy}}$,~~$\frac{e_{x}(l_{y}^{2}-e_{y}^{2})-e_{y}(l_{x}^{2}-e_{x}^{2})}{l_{y}(l_{x}^{2}-e_{x}^{2})\sin\theta_{xy}}$
}\\
$\bar{X}_{z},\bar{Y}_{z}$&\multicolumn{2}{c}{$X_{zx}^{2}+X_{zy}^{2}-2X_{zx}X_{zy}\cos \phi_{z}$,~~$Y_{zx}^{2}+Y_{zy}^{2}-2Y_{zx}Y_{zy}\cos \phi_{z}$}\\
$\bar{Z}_{z}$&\multicolumn{2}{c}{$X_{zx}Y_{zx}+X_{zy}Y_{zy}-(X_{zx}Y_{zy}+Y_{zx}X_{zy})\cos
\phi_{z}$}\\
$\alpha_{x},\beta_{z}$&\multicolumn{2}{c}{$\arccos\left(\frac{-X_{xy}Y_{xy}+\sqrt{1+X_{xy}^{2}-Y_{xy}^{2}}}{1+X_{xy}^{2}}\right)$, 
$\arccos\left(\frac{-\bar{Z}_{z}+\sqrt{\bar{Z}_{z}^{2}-\big(\bar{X}_{z}+\sin^{2}\phi_{z}\big)\big(\bar{Y}_{z}-\sin^{2}\phi_{z}\big)}}{\bar{X}_{z}+\sin^{2}\phi_{z}}\right)$}\\
$\gamma_{xy},\Gamma_{z}$&\multicolumn{2}{c}{$\arccos\left(\frac{X_{xy}\cos\beta_{x}+Y_{xy}}{\sin\beta_{x}}\right)$,~~$\sin\beta_{x}\sin\gamma_{xy}$}\\
\end{tabular}
\end{table}
\indent In summary, we provided a complete analysis to minimum error discrimination of four mixed qubit states with arbitrary prior probabilities.
For minimum error discrimination of four mixed qubit states with arbitrary prior probabilities, we analyzed the
necessary and sufficient conditions for the existence of null measurement operator. Specifically, from the geometric structure of qubit states, we obtained the analytic
condition for deciding the existence of a null operator in
minimum-error measurement for four qubit states, which lets us figure out the 
necessary and sufficient conditions for every optimal POVM to have non-zero elements. 
From the condition, we completely analyzed minimum-error discrimination of four
qubit states with arbitrary prior probabilities. In addition, we provided a relevant example.\\
\section*{Acknowledgements }
\indent When we were writing this manuscript, we got to know the preprint(https://arxiv.org/abs/2108.12299), which deals with a similar subject. We thank Prof.Bae Joonwoo for valuable discussion and contribution to writing an early version of manuscript.
This work is supported by the Basic Science Research Program through the National Research Foundation of Korea (NRF) funded by the Ministry of Education, Science and Technology (NRF2018R1D1A1B07049420) and Institute of Information and Communications Technology Planning and Evaluation (IITP) grant funded by the Korean government (MSIT) (No. 2020001343, Artificial Intelligence Convergence Research Center (Hanyang University ERICA)). D.H. acknowledges support from the National Research Foundation of Korea (NRF) grant funded by the Korean government (Ministry of Science and ICT) (NRF2020M3E4A1080088).

\end{document}